# Reliability-Based Fault Analysis and Modeling of Satellite Electrical Power Subsystems Using Fault Tree and Simulation Tools

1- Niloofar Nobahari, 2- Alireza Rezaee

1, PhD Student, Department of Mechanical Engineering, Old Dominion University, Norfolk, VA, USA

: Nnoba001@odu.edu

2Associated professor, Faculty of New sciences and technologies, University of Tehran, Tehran, Iran

E-mail: arrezaeee@ut.ac.ir

One of the most important satellite's subsystems is its electric power subsystem. The occurrence of a fault in the satellite power system causes the failure of all or part of the satellite. Calculating the overall reliability of the power system before the mission plays a crucial role in improving the design of the satellite power system. Each component of the power system may malfunction due to pressure, launch pressure, and operating conditions. Accordingly, in this paper, first a healthy and faulty system for the components of the electrical power system is simulated with MATLAB. Finally, by drawing a fault tree to analyze the reliability of the power subsystem, overall mission reliability, power system fault rate and overall fault rate of the mission is calculated by Windchill software. Finally, the total mission assurance of 0.999 was obtained, which indicates the high reliability of the simulated system.

**Introduction**

According to the mission, satellites have the necessary parts and subsystems to meet the needs [1]. Electricity generation, management and distribution subsystems are important part of the electrical systems of all satellites. According to the type of mission and its requirements and needs, this system consists of subsystems such as photovoltaics to convert solar energy and generate electrical power, batteries, and the bus of the electrical aggregation unit, which each these subsystems are designed and implemented for a specific purpose in the system. Each component of this system may be due to tolerance Stresses, launch pressure, and working conditions which can make different faults and malfunctions [2]. Using the probability of any fault occurring in the battery, bus, solar arrays and power distribution subsystems, which the most faults occur in these parts and sabotage the whole mission and leads it to fail help to simulate and design the system without any errors and faults. Make and construct the satellite is time and money consuming and it is very difficult and sometime impossible to repair any part of it if any fault occurs in any parts. It seems it is mandatory to recognize possible faults, especially in electrical power system, to prevent the mission failure and continue it, as well, in the design stage before it cause serious damage in other related subsystems. A fault in the electrical power system of satellite may cause the satellite to fail completely. The satellite designer should be vigilant about the most important electrical parts which most faults happen in it [3]. The most important subsystems which are common in every satellite are attitude determination control, communication, control and data handling, science equipment, structure, and electrical power. The electrical power system is equipped with different subsystems such as photovoltaic arrays, solar cells, bus, batteries, and wires. [4].

**Fault in electrical power system of satellite**

If the electrical power system breaks down repeatedly over time and the problem would not fix properly in time, it will cause the whole system to fail. Different malfunctions due to the space environment in the satellite may occur in the space anytime. Combination between EPS fault and other subsystem's faults most of the time walk hand in hand with each other. Common failures in the electrical power system of satellite include mechanical failure, wire failure, short circuit, solar cell failure, battery failure, computer failure and collision failure. Is space, combining each of these failures together can occur individually or combine with other failures. To calculate the probability of the worst case that causes a fault in the electrical power system of satellite, there is a risk estimation method. In a more complex system such as electrical power, it is not possible to check all faulty states without analyzing all possible faults in the system. All faults that may occur in the electrical power system based on their output current for each fault, and its influence on the other related subsystems are calculated and analyzed. Probability Analysis Framework for this analysis [5] a network with several sources such as power system is given. This method is used to measure the ability of the system through the system topology and simulation of several faults. System topology refers to the number of components, the type and how they relate to each other. To investigate the occurrence of several faults in the electrical power system of the satellite, the LEO satellite is studied and the current in it is investigated. The electrical power diagram used in Figure 1 is given. This paper is analyzing most three important components which have the highest probability of fault occurrence: battery, solar panel and electrical distribution unit (Figure 1). Research on the detection and classification of these faults has been conducted in the electrical equipment. In the quasi-monitoring graph learning method, faults that are not able to be detected in photovoltaic systems are identified. For fault detection in converters, a new topology for fault tolerance, which is very fast reconfiguration fault detection, and for experimental testing, control and fault detection, configuration schemes on an FPGA gate chip feature are used. In the comparative method of fault detection in the battery, for accurate expression of the battery status as well as checking the discharge fault in it from several models such as recursive least squares estimator (RLS), non-linear equivalent battery circuit (grade 3 model including resistor, capacitor and Voltage source) and extended Kalman filter are used [6]. But these studies do not cover the entire electrical power system of a satellite and faults. Only in two scientific papers in faults realm have been detected in the power system of small satellites. In the first study, faults detect only in three parts: voltage, current and temperature sensors based on PCA method. Therefore, the other most important component which the other most important faults may occur was not investigated. On the other hand, this paper only reviews micro-satellites based on PCA method for various types of faults in voltage, current and temperature sensors. This method is useful only for recognize one fault in the system and if two or more faults occur in the system simultaneously this method is not work anymore. According to previous articles, all faults in the EPS that have been detected in the of photovoltaics, electrical distribution unit and batteries, and the probability of each fault occurring are calculated and these data are entered in the Windchill software. The output of the software is fault three, risk matrix of the simulated EPS. The following fault tree Draw the electrical power system of satellite. Therefore, in this study, it is tried that the most faults that can occur in this system have been calculated by drawing the fault tree and the overall reliability of the power subsystem [7].

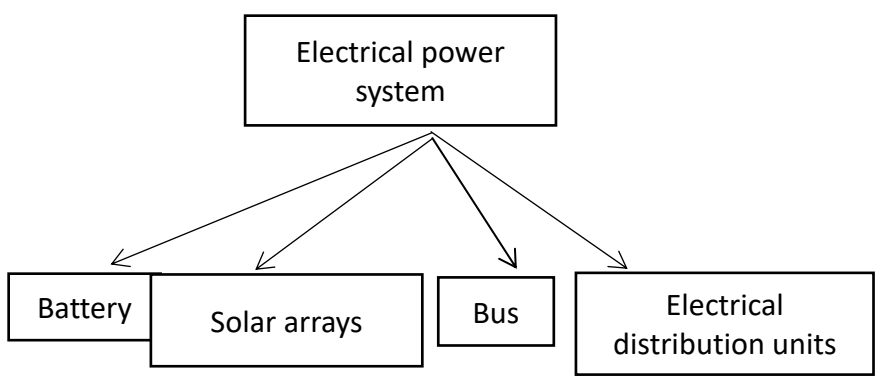

Figure 1 Diagram of the connection of the components of the electrical power subsystem of satellite [7]

## Fault-free and faulty system modeling

In order to evaluate and analyze the performance of the satellite electrical power system, first a healthy system (without fault) should be modeled to compare its performance with the faulty system to identify the fault and fault of the system. Batteries, solar panels and bass were simulated safely and in case of faults with MATLAB software, then the overall reliability of the simulated power system was calculated with Windchill software.

## Battery simulation for healthy and faulty states

In order to choose the most suitable battery for a space project, we need to look at our needs and parameters such as: orbital position and deflection angle. Satellite orbit, temperature effects, amount of stored energy, satellite mass and type of mission, charge efficiency and most importantly to withstand the conditions and stresses of launch in space. However, given the advances in science in this field, it can be said that by considering all the parameters raised and comparing the performance of different types of batteries, for a hypothetical mission in the orbit of LEO and cubic satellite with Nano-dimensions use Li-Ion batteries will be useful in many ways. Therefore, to design and extract the desired battery specifications, we must first calculate the number of cells used in the battery according to the line voltage based on equation (1). The average discharge voltage of Li-Ion cells is approximately equal to 3.6 volts [8].

$$V_{line} = N.V_{cell} \qquad (1)$$

In the above relation, $V_{line}$ is the line voltage, N is the number of cells and $V_{cell}$ is the average voltage of each cell. Taking into account, the average lithium ion cell voltage and line voltage, the number of battery cells is 5. Now, according to the length of the mission (2 years) and the number of charge and discharge cycles of the battery, the desired discharge depth and determined with appropriate margin of confidence after determining the discharge depth, equation (2) can be used to calculate the desire battery capacity.

$$c = \frac{P_e.T_e}{N_b.\eta_{dis}.[(N-1)V_{cdis} - V_d].DOD} \qquad (2)$$

In this regard, C is the battery capacity in ampere-hours, $P_e$ is the required charge during the shadow phase, $T_e$ is the shadow phase duration in hours, $N_b$ is the number of parallel batteries, $\eta_{dis}$ discharge converter efficiency, $V_{cdis}$ average discharge voltage in volts And DOD is the discharge depth of the battery. Nano-satellites often use one battery, so the value of $N_b$ is 1. After specifying the above parameters, the battery capacity is calculated to be 2.3 ampere-hours. Then, for modeling the battery according to the discharge curve of Li-Ion cell, the values of nominal voltage, rated capacity, maximum capacity, full charge voltage, rated discharge current, capacity at rated voltage and exponential region are determined [8]. Based on the mentioned information and cases, a healthy and fault-prone one has been implemented in MATLAB. The faults have been applied to the system artificially. figure 2.

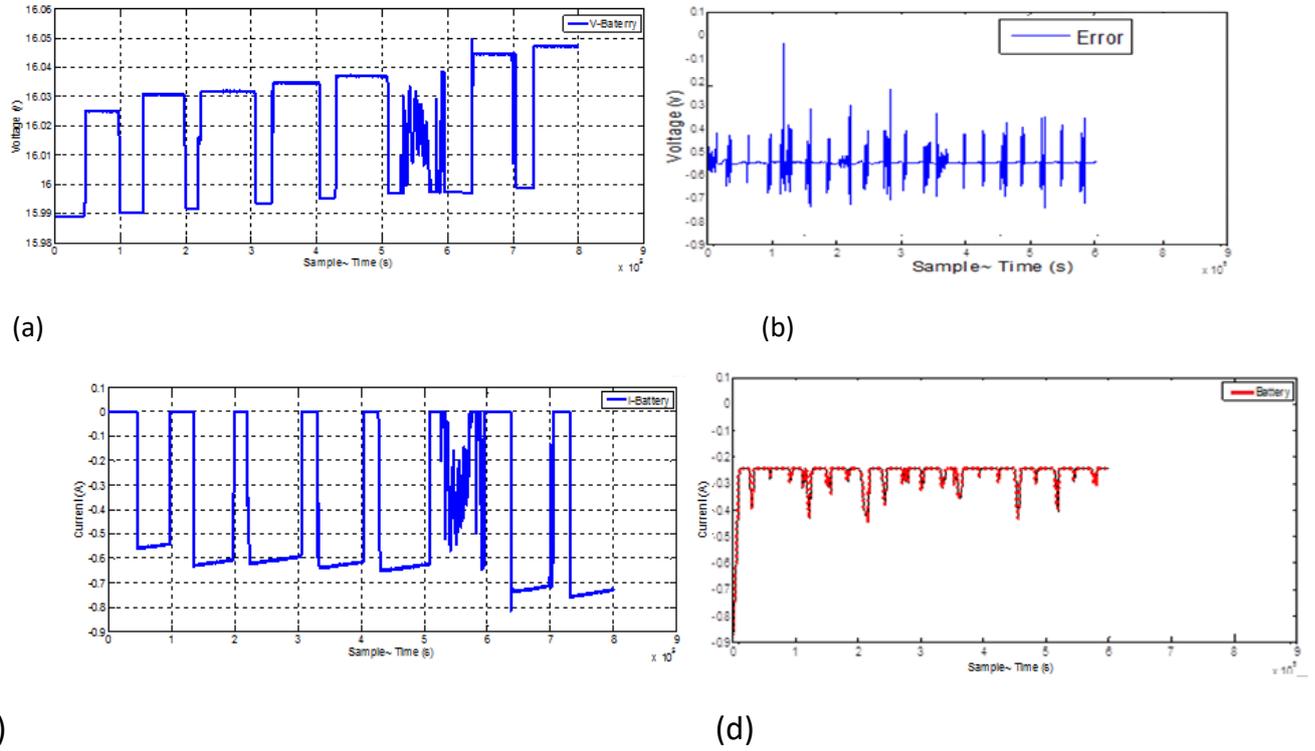

(a) (b)

(c) (d)

Figure 2 Simulated healthy and faulty diagram for battery. (a) Voltage-time for healthy battery (b) Voltage-time for faulthy battery (c) Current-time for healthy battery (d) Current time for faulthy battery.

**Solar arrays simulation for healthy and faulty states**

Solar energy is one of the most important sources of energy supply in space systems. This energy is converted into electricity in various ways. The use of solar cells in the direct conversion of solar energy into electricity is one of these methods. Conventional faults in solar arrays are found in open circuit fault, grounding fault, line fault (line to line) and mismatch fault in photovoltaic modules [9].

The production of the photovoltaic section as well as the performance of the whole system is affected by the orbital conditions and environmental factors of the satellite such as orbital height, inclination angle, radiation intensity, heat transformation, and radiation in the orbit and temperature. Proper execution of the mission requires the correct operation of all systems and components of the satellite, but due to the stresses and pressures that when launching the satellite into orbit, as well as its operating conditions, faults may occur in all parts of the satellite, including system Its electrical power occurs. If these faults are not detected and fixed in time, they may lead to system disruption and even cancellation of the mission. Therefore, these faults should be identified by using some methods.To calculate the number of series and parallel solar cells, it must be first calculated the electrical power required by the satellite, including the suitable reliability margin and current line based on equation (3) and (4).

$$P_{sa} = \frac{1}{\eta_{pc}.\eta_d}(P_{av} + \frac{P_p.t_p}{t_s}\frac{V_{max}}{V_{min}}) \tag{3}$$

$$I_{bus} = \frac{P_{sa}}{V_{bus}} \tag{4}$$

According to equation (4) $P_{sa}$ required electric power of the satellite, $\eta_{pc}$ electrical energy conversion efficiency, $\eta_d$ deviation following the maximum power, $P_{av}$ average power consumption of the satellite, $P_p$ power The maximum consumption during the shadow phase is $t_p$ and $t_s$, respectively, the maximum power demand time and the solar radiation time, and $V_{min}$ and $V_{max}$ are the minimum and maximum values of the system battery voltage. And based on equation (4-6) $I_{bus}$ are the current and $V_{bus}$ are the line voltage.

Now, after calculating the required satellite power and line current, the number of series cells and the number of parallel filaments can be determined using Equations (5) and (6), respectively [10].

$$N_s = \frac{V_{bus}}{V_{mp(EOL)}} \tag{5}$$

and

$$N_{ps} = \frac{I_{bus}}{I_{mp(EOL)}} \tag{6}$$

Common types of faults considered in the simulation include grounding fault, line fault, and mismatch fault. Grounding fault is a random fault that is a short circuit of one or more components of the circuit that have an electric current to the ground circuit. Line fault is also a random fault and a short circuit between two points that have a potential difference in the array They are solar, they happen. Nonconforming fault occurs when some electrical parameters of the solar module change significantly compared to other modules. This fault can be temporary as a partial shadow on solar arrays or permanent as an open circuit due to wear over time [11]. Figure 3 shows the simulation of healthy and faulty solar arrays.

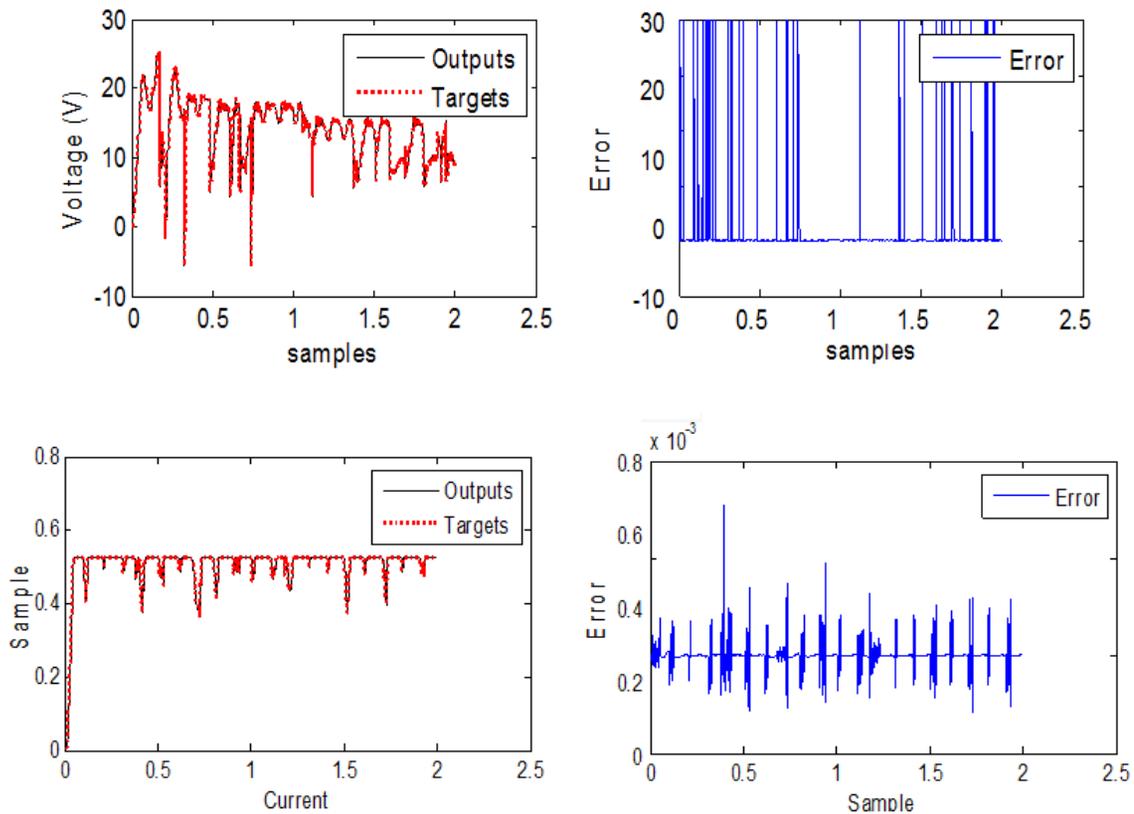

Figure 3 solar array diagram in the faulty and healthy state. (a) Voltage-time diagram for healthy solar array (b) Voltage-time diagram for faulty solar array (c) current-time diagram for healthy solar array (d) current-time diagram for faulty solar array

**Fault tree analysis as a fault diagnostic tool**

Troubleshooting and rooting methods can be divided into two general categories based on model and data-based, each of which is divided into two sub-categories of quantitative and qualitative methods. Reference [10] categorized these methods in 2008 and considered the defect tree as one of the cause and effect models that is part of the methods based on the qualitative model. The reason for the quality of this method is that it must be done by an expert for each modeling system and it can be said that this method is based on the structure of the system. One of the methods of analysis is the top-down and event-oriented approach. This method of analysis is based on a series of logic rules and follows the laws of Boolean algebra and leads to the production of a logical model of logic gates. To analyze a failure state of the system, fault tree analysis can be performed. In this analysis, through Boolean algebra, we can combine a set of low-level events as a class to reach the main failure state. Important features of this analysis are:

1. Better understanding of the logic of the failure event

2. Control and monitor the process of reliable operation in complex systems

3. One of the design tools to help find the important requirements of the system

In this method, the failure state is considered as the highest event and the conditions that lead to the failure state event are added to a lower level through logic gates. In the same way, we add the conditions and reasons that lead to each of the reasons for the occurrence of the main failure to a lower level. This process continues until we reach the basic reasons [12]. Figure 4 shows the process for forming a fault tree for a satellite electrical power system.

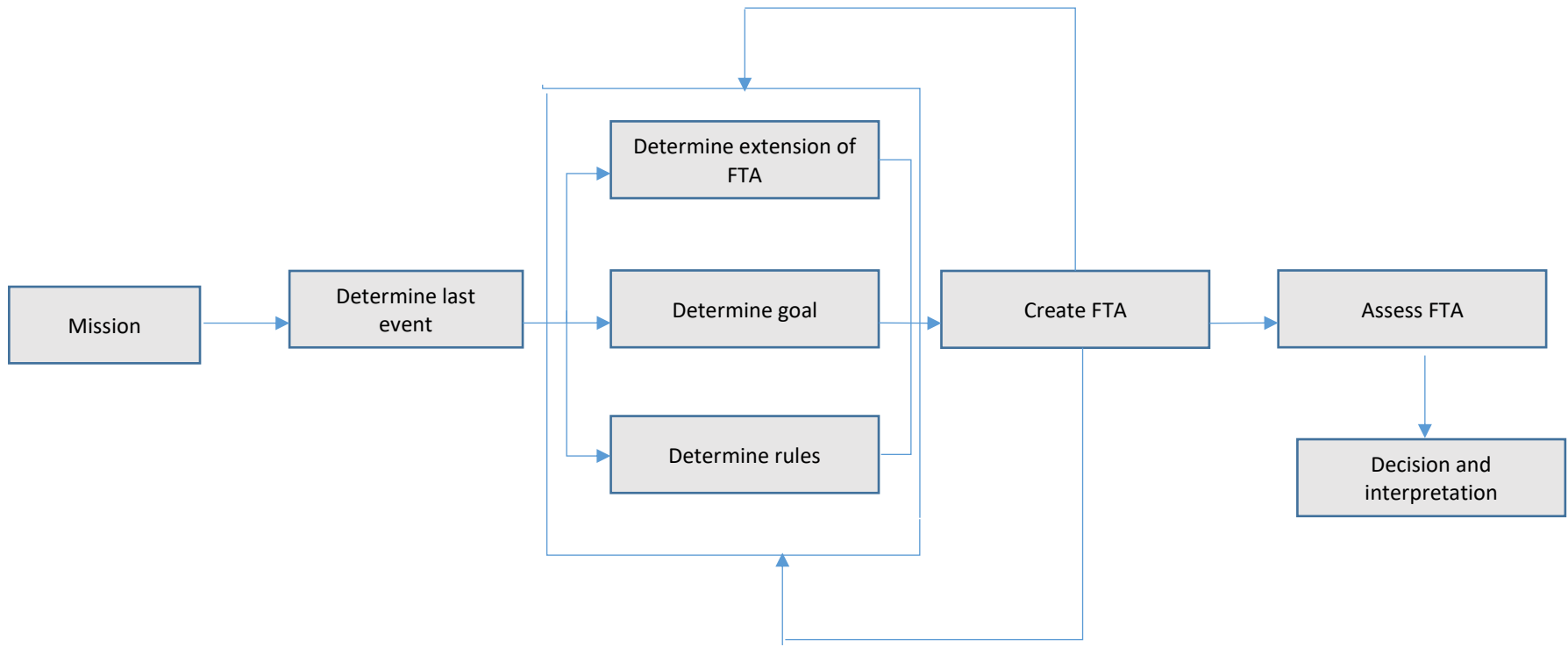

Figure 4 Fault tree formation steps for satellite electrical power subsystem

**Fault Tree Diagram Symbols**

There are two basic types of fault tree diagram notations: events and logic gates. The primary or basic failure event is usually denoted with a circle. An external event is usually depicted with a symbol that looks like a house. It's an event that is normal and guaranteed or expected to occur. Undeveloped event usually denotes something that needs no further breakdown or investigation or an event for which no further analysis is possible because of a lack of information. A conditioning event is a restriction on a logic gate in the diagram. These gate symbols describe the Boolean relationship between outcomes. Gate symbols can be the following:

- OR gate - An event occurs as long as at least one of the input events takes place
- AND gate - An event occurs only if all input conditions are met
- Exclusive OR gate - An event occurs only if one of the input conditions is met, not if all conditions are met
- Priority AND gate - This is probably the most restrictive scenario when an event occurs only after a specific sequence of conditions

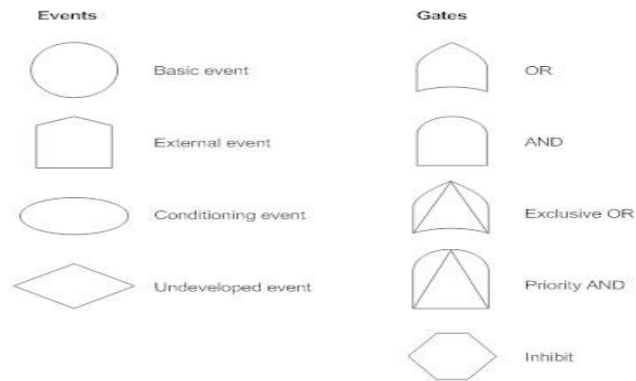

Figure 5 fault tree symbols

**Fault tree formation and drawing for satellite electrical power system**

Most electrical systems, including satellite electrical power systems, consist of electrical and electronic equipment such as analog and digital circuits that include transistors, Thyristor, diodes, amplifiers, logic elements, switches and cables, etc. Each of the electrical and mechanical components made by the manufacturers has a certain reliability that can be expressed using other interpretations. After calculating the reliability coefficient, another parameter called the fault rate in equipment (λ) can be defined according to Equation (7) as follows [13]:

$$\lambda = \frac{1}{t_{int}} \times \frac{N_f}{N_t} \qquad (7)$$

In this regard, $t_{int}$ is the time to check the performance of the equipment, $N_f$ and $N_t$ are the number of faulty elements and the total elements, respectively. The parameter λ depends on the temperature and operating modes of the system. Different work Table (1) shows the values of λ of a number of electrical and electronic components. The fault rate in the table for calculating the reliability of the electrical power system of satellite is entered in the fault tree to calculate the overall reliability of the system.

Table 1: Fault rate at 40°C temperature of some electrical and electronic equipment used in satellite electrical power system [20].

| $\lambda(h^{-1})$ | Fault rate in electrical and electronic equipment |
|---|---|
| $10^{-9} \times 70^{-1}$ | Transistor |
| $10^{-9} \times 360^{-36}$ | Thyristor |
| $10^{-9} \times 30$ | Digital integrated circuits |
| $10^{-9} \times 30$ | Logical elements |
| $10^{-9} \times 2000$ | Analogue switch |
| $10^{-9} \times 900^{-300}$ | Amplifier |
| $10^{-9} \times 6^{-1}$ | Diodes |
| $10^{-9} \times 300^{-200}$ | li-Ion Battery |
| $10^{-9} \times 200^{-100}$ | Solar arrays |

After drawing the fault tree, marking the events and writing the vertex event equation is done in terms of minimum segments.

Figure 5 shows the fault tree of the electrical power system of satellite using the reliability blocks in the Windchill software.

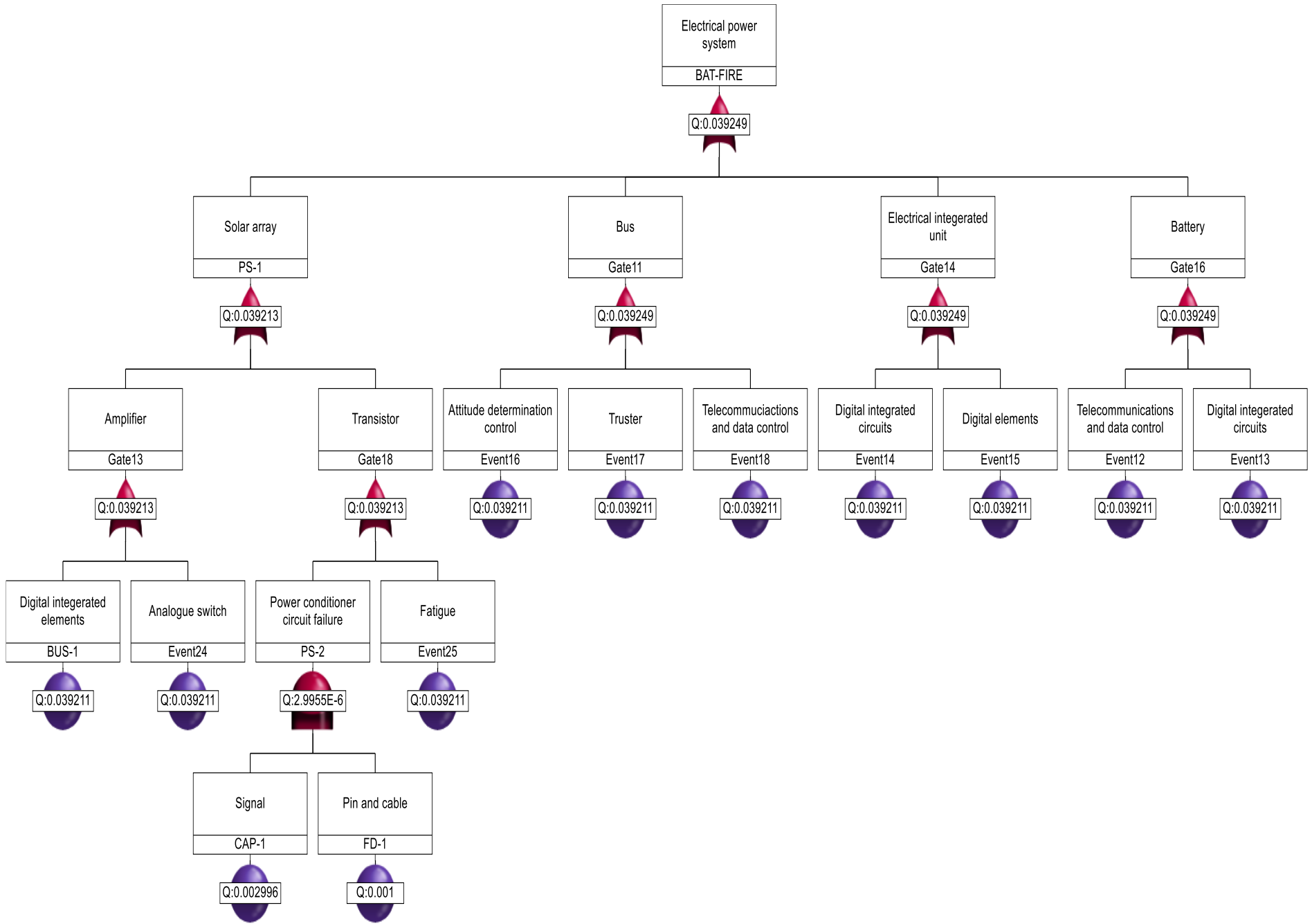

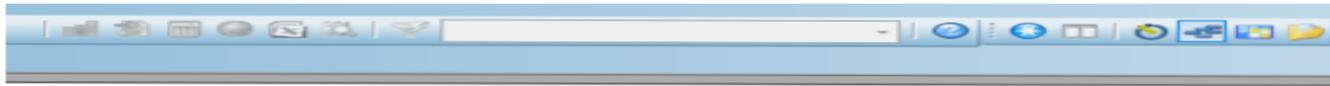

| Name | Description | Gate/Event Type | Input Type | FR/MTBF |
|---|---|---|---|---|
| BAT-FIRE | Electrical power system | OR Gate | | |
| BUS-1 | Digital integrated elements | Basic Event | Failure with Repair | Failure Rate |
| CAP-1 | Signal | Basic Event | FR/MTBF | Failure Rate |
| Event12 | Telecommunications and data control | Basic Event | Failure with Repair | Failure Rate |
| Event13 | Digital integrated circuits | Basic Event | Failure with Repair | Failure Rate |
| Event14 | Digital integrated circuits | Basic Event | Failure with Repair | Failure Rate |
| Event15 | Digital elements | Basic Event | Failure with Repair | Failure Rate |
| Event16 | Attitude determination control | Basic Event | Failure with Repair | Failure Rate |
| Event17 | Truster | Basic Event | Failure with Repair | Failure Rate |
| Event18 | Telecommuciactions and data control | Basic Event | Failure with Repair | Failure Rate |
| Event24 | Analogue switch | Basic Event | Failure with Repair | Failure Rate |
| Event25 | Fatigue | Basic Event | Failure with Repair | Failure Rate |
| FD-1 | Pin and cable | Basic Event | Constant Probability | |
| Gate11 | Bus | OR Gate | | |
| Gate13 | Amplifier | OR Gate | | |
| Gate14 | Electrical integrated unit | OR Gate | | |
| Gate16 | Battery | OR Gate | | |
| Gate18 | Transistor | OR Gate | | |
| PS-1 | Solar array | OR Gate | | |
| PS-2 | Power conditioner circuit failure | AND Gate | | |

File Name: nobahari.rfp
Top Gate: BAT-FIRE
No. of Gates: 8
No. of Events: 12

Page #:3

Figure 6 Fault tree graph and result for satellite electrical power system failure in Windchill software

Finally, in Figure 7 of the frequency-power diagram, the electrical power of the satellite was simulated as healthy and faulty states. The red lines indicate that the subsystem is faulty state and the black lines indicate that the subsystem is in healthy state.

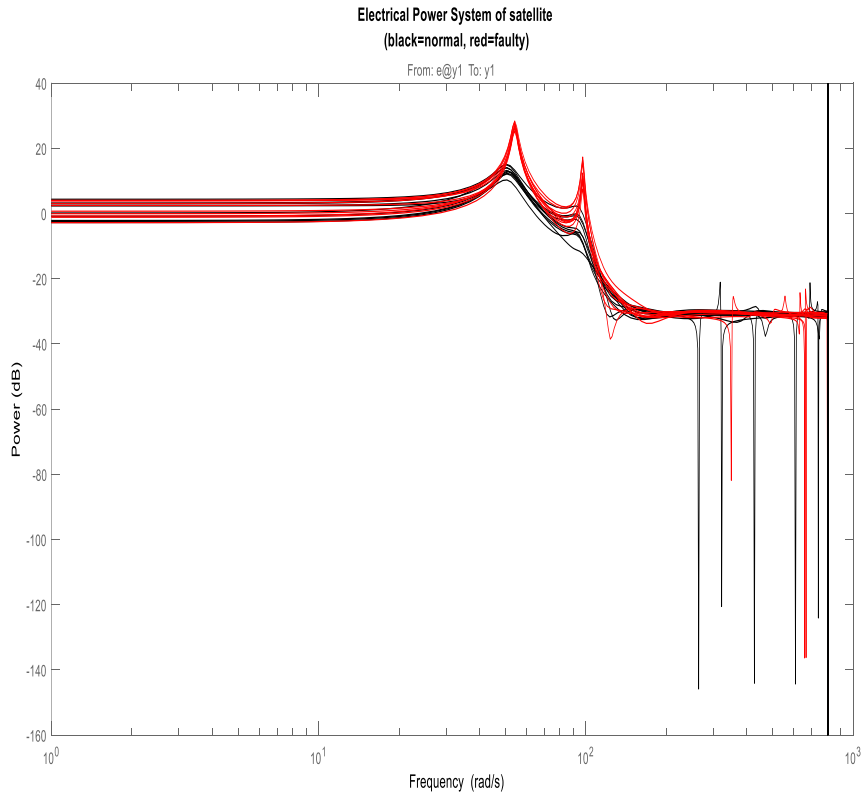

Figure 7 Frequency diagram - Faulty and healthy satellite power system power

**Computational analysis**

Under the electrical power subsystem and the computational algorithm used to draw the fault tree, it is assumed that all faults lead to the failure of the power subsystem. A fault is also plotted and calculated for each event. It is also assumed that at least one fault occurs in the system and at least one power source is available. Has been. In a system with M element number, N = 2M, there is a possible fault scenario and F specifies the number of fault scenarios that cause the power system to malfunction [14].

For the drawn fault tree, 11 events have been considered, the number of fault scenarios for the power system is approximately 2048 combined faults. The probability of complete failure of the power system P (S) is obtained from Equation (8). Also, the probability of power system response in case of m fault is obtained from Equation (9).

P(S)=F/N

$P_m(S) = S~(m) / N~(m)$ , $P_m(R) = R~(m) / N~(m)$ , $P_m(F) = F~(m) / N~(m)$ (8)

Where $\sum_{m=1,\ldots,M}^{M} N(m) = N$ , $\sum_{m=1,\ldots,M}^{M} R(m) = R$ , $\sum_{m=1,\ldots,M}^{M} S(m) = S$ , $\sum_{m=1,\ldots,M}^{M} F(m) = F$ 9)

The probability of the power system response is obtained from the sum of relations (9). The probability considered for the occurrence of each fault is used [15] then by placing the system response probability in the fault tree event block, the overall reliability of the power system, mission reliability, fault rate, mission availability, system reliability as Software output is obtained. The overall output of system reliability over time is shown in Figure 7 [16].

The table below shows the output of the generated fault tree.

| Value | Result |
|---|---|
| Failure Rate, Predicted | 23.117593 |
| Reliability, Predicted | 0.99810 |
| Availability | 0.999997 |
| Failure Rate, Mission | 27.368133 |
| Reliability, Mission | 0.439972 |
| Availability, Mission | 0.999997 |

Figure 8 Table of output results of electrical power satellite from Windchill software of the overall reliability.

**Risk matrix**

The output of the risk matrix has three colors: green, yellow and red. Low risk is shown in green, medium risk in yellow and high risk in red. The final risk matrix for the electrical power system of satellite is shown in Figure 8. The numbers written inside the risk table indicate the low risk of the designed system. [18-20]

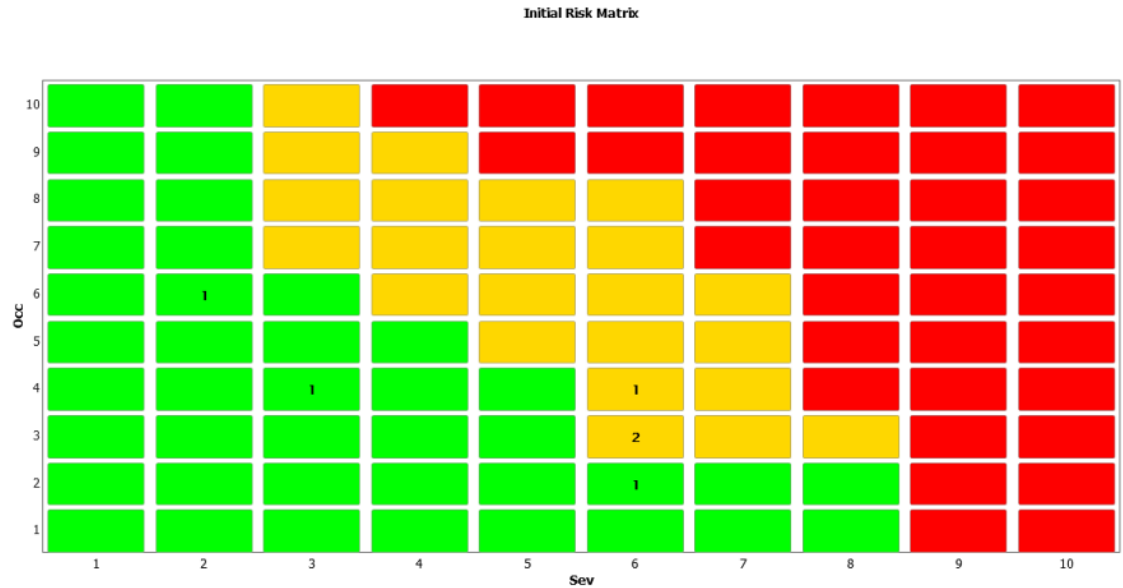

Figure 9 Satellite electrical power system risk matrix

Due to the fact that the failure in the power source cause the failure in the whole system and leads the mission fails at the end. Therefore, the electrical power system of satellite has a vital role. Diagnosis faults and causes of faults help to troubleshoot and properly design the satellite's electrical power subsystem. In this article first, a healthy electrical power system is simulated, then by artificially entering the fault into the healthy system and faulty those faults of the battery system, solar arrays and the distribution of electrical power were simulated with MATLAB software. Then, the probable faults that are most likely to occur and cause the complete failure of the electrical power system and ultimately the failure of the mission have been identified. These faults, as events, are the entries of the fault tree in Windchill software. Using the designed fault tree with the overall reliability of the satellite mission (0.998) and the reliability of the satellite power system (0.999). According to the reliability diagram of Figure 9, it is clear that the calculated reliability for the designed power system indicates the high reliability of the designed system.

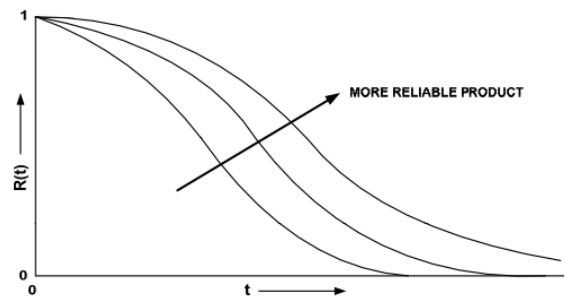

Figure 10 Overview of Reliability [20].